\documentclass[%
preprint,
 amsmath,amssymb,
 aps, physrev,
]{revtex4-2}

\usepackage{graphicx}
\usepackage{dcolumn}
\usepackage{bm}
\usepackage{amsmath}   
\usepackage{enumerate}
\usepackage{braket}

\begin{document}

\title{Universal and Tunable Sudden Freezing of Entanglement Volume} 

\author{Luchang Niu}
\affiliation{Department of Physics and Astronomy, University of Rochester, Rochester, New York 14627, USA}
\email{lniu5@u.rochester.edu}

\author{Joseph H. Eberly}
\affiliation{Center for Coherence and Quantum Optics, and Department of Physics and Astronomy, University of Rochester, Rochester, New York 14627, USA}

\date{\today}

\begin{abstract}
In a system where two identical two-level atoms interact with their common one-mode cavity field, it is shown that entanglement can become abruptly frozen in time, remaining at a constant value for a period of time until it begins to thaw from this value from the entanglement sharing perspective [Ding \textit{et al.}, \textit{Phys. Rev. A} \textbf{103}, 032418 (2021)]. We generalize this exotic behavior of entanglement sharing dynamics to more general systems with arbitrary N qubits, instead of restricting to the atom-cavity mode interaction system. We also demonstrate methods to control the entanglement freezing time and freezing value, and we discover a nontrivial dynamics where entanglement is frozen permanently. In addition, we show that this phenomenon is not a coincidence but a universal feature in a variety of systems with a geometric explanation of the mechanisms.
\end{abstract}

\maketitle

\section{Introduction}
Quantum entanglement, a unique non-classical form of correlation, has become the central principle of many modern quantum technologies such as quantum communication, computation, and metrology, providing advantages and supremacy over classical means \cite{horodecki2009quantum}. Recently, quantum entanglement has been widely recognized as a fundamental resource that distinguishes quantum mechanics from classical physics \cite{nielsen2010quantum,chitambar2019quantum,bennett1996mixed}. Its resource nature has led to many advancements in quantum information theory and quantum information processing applications that treat entanglement as a quantifiable currency for many tasks. The importance of quantum control has been increasingly recognized since practical quantum technology requires very robust manipulation of entangled states \cite{wiseman2009quantum,dong2010quantum}. Understanding the mechanism underlying entanglement dynamics that enable or hinder entanglement control has become one of the central themes in quantum information science \cite{rivas2012open,yu2009sudden}.

One of the most intriguing dynamics of entanglement is entanglement freezing. Bipartite entanglement \cite{chanda2018scale}, Quantum discord (a different form of quantum correlation) \cite{mazzola2010sudden,haikka2013non}, and multipartite entanglement \cite{ali2017freezing,ali2014sudden,wu2016frozen} have shown both theoretically \cite{aaronson2013comparative,cianciaruso2015universal} and experimentally \cite{liu2016time,xu2010experimental} to become frozen in certain open quantum systems under decoherence. Recently, people have demonstrated pure-state entanglement freezing from the entanglement sharing perspective in a particular lossless system, which proves that decoherence is not necessary for the freezing effect \cite{ding2021sudden,qian2021freezing}. 

A natural question then arises: Do different entanglement freezing observations have a universal mechanism or merely a coincidence? A few studies have proved the general existence of entanglement freezing in a specific type of open quantum system \cite{aaronson2013comparative,cianciaruso2015universal}, but without identifying the underlying physical channels or mechanisms. By contrast, most reported entanglement freezing observations only occur in some specific systems with carefully chosen parameters and initial states. But it remains unclear why entanglement freezing happens in the first place \cite{ding2021sudden}. In other words, no universal explanation for the origin of entanglement freezing has been given for any current observations.

In this work, we uncover a class of entanglement freezing dynamics in general N-partite systems with a universal mechanism in multiple lossless systems. The universality of this phenomenon is reflected on its appearance in many different physical systems, independent of their microscopic details. We show that such entanglement freezing is tunable: both frozen value and freezing duration can be independently controlled, and permanent freezing with tunable values is allowed for suitably chosen initial states. Finally, we identify and analyze the fundamental mechanism of this effect with a broad applicability that provides a universal explanation for different classes of entanglement freezing previously reported in different contexts.

This paper is organized as follows: In Sec.~\ref{section2}, we introduce the concept of ``entanglement volume", in which freezing is observed. Sec.~\ref{section3} presents two representative examples - one in a closed system and the other in an open system - to illustrate the universality of the phenomenon. In Sec.~\ref{section4}, we provide an algebraic explanation of the freezing dynamics, and give a universal and graphical interpretation that applies to many other forms of entanglement freezing in Sec.~\ref{section5}. Finally, Sec.~\ref{section6} summarizes our results and conclusions. 

\section{Entanglement Volume}
\label{section2}
To investigate entanglement dynamics, we focus on the time evolution of the entanglement volume~\cite{ding2021sudden}, introduced from the perspective of entanglement sharing. The entanglement volume is defined as the sum of all one-to-other bipartite entanglements based on the normalized Schmidt weight~\cite{qian2018entanglement}. The normalized Schmidt weight $Y$ is given by $Y = 1 - \sqrt{2/K - 1}$, where $K$ is the standard Schmidt weight. It is also related to the concurrence~\cite{wootters1998entanglement} through $Y = 1 - \sqrt{1 - C^2}$, where $C$ denotes the bipartite concurrence. Originally, concurrence was proposed to quantify entanglement in a two-qubit system and was later generalized to bipartite systems of arbitrary dimension~\cite{rungta2001universal}. For any pure state $\ket{\psi} \in \mathcal{H}_A \otimes \mathcal{H}_B$ with dimension $D_1\times D_2$, the concurrence is defined as $C(\psi) = \sqrt{\frac{d}{2(d-1)}(1 - \mathrm{Tr}\,\rho_A^2)}$, where $d=\text{min}\{D_1,D_2\}$, and $\rho_A = \mathrm{Tr}_B(\ket{\psi}\bra{\psi})$ is the reduced density matrix of subsystem $A$. For a system containing more than two parties, one can define a one-to-other concurrence as the bipartite entanglement between a single qubit and the remainder of the system~\cite{coffman2000distributed}. Denoting this as $C_{A_i|\bar{A}_i}$, the corresponding normalized Schmidt weight is 
\begin{equation}
Y_{A_i|\bar{A}_i} = 1 - \sqrt{1 - C_{A_i|\bar{A}_i}^2}, \qquad i = 1,2,\dots,N,
\end{equation}
where $A_i$ denotes the $i$th qubit and $\bar{A}_i$ represents the remaining $N-1$ qubits. The total entanglement volume of an $N$-qubit pure state is then defined as 
\begin{equation}
Y_s = \sum_{i=1}^{N} Y_{A_i|\bar{A}_i},
\label{Eq:entanglement volume}
\end{equation}
which was first studied for the $N=3$ case in Ref.~\cite{ding2021sudden} and is here extended to arbitrary $N$. Under a given Hamiltonian, the state evolves as $\ket{\psi_t} = e^{-iHt}\ket{\psi_0}$, making $Y_s$ a time-dependent quantity $Y_s(t)$.

The concept of entanglement sharing~\cite{ding2021sudden} emphasizes that each one-to-other bipartite entanglement captures only part of the total entanglement in the system, and the total volume $Y_s$ is shared among them. During evolution, the entanglement associated with each one-to-other bipartition can redistribute, sharing the entanglement volume which is also evolving with time. For an $N$-qubit system, each $Y_{A_i|\bar{A}_i} \in [0,1]$, so the maximum possible entanglement volume is $Y_{s,\mathrm{max}} = N$. In the following sections, we show that this shared volume can exhibit sudden freezing and thawing during the system’s evolution. This entanglement-sharing perspective provides new insights for understanding and controlling entanglement dynamics.

\section{Sudden Freezing and Thawing of Entanglement}
\label{section3}
In this section, we demonstrate entanglement sudden freezing in specific models with very distinct physical settings to illustrate the universality of the phenomenon: a closed three-qubit system described by the quantum XX model and an open quantum system of cavities coupled to reservoirs. Both cases exhibit sudden freezing and thawing of entanglement volume.

\subsection{The Quantum XX Model}
To connect our analysis to experimentally relevant platforms, we consider ultracold atoms confined in an optical lattice generated by multiple standing-wave laser beams. In the Mott-insulating regime, where tunneling is strongly suppressed, each lattice site is occupied by at most one atom. Each atom can be treated as an effective two-level system with the two internal hyperfine states identified as pseudo-spin states $\ket{\uparrow}$ and $\ket{\downarrow}$. Such spin models have been extensively studied and known to provide controllable quantum simulators for a wide range of quantum phenomena \cite{duan2003controlling,bloch2008many,bloch2012quantum}. Importantly, this mapping allows the entanglement dynamics we describe below can, in principle, to be observed in real experimental setups.

Following the approach in \cite{duan2003controlling,qian2021freezing}, we assume all atoms are aligned in a 1D chain and each atom interacts only with its two nearest neighbors. This configuration is known as an optical lattice chain. The system's Hamiltonian is given by
\begin{equation}
H=-\sum_{k, s}(t_{\mu,s}a_{k,s}^\dagger a_{k+1,s}+\text{H.c.})+\frac{1}{2}\sum_{k,s}U_s(n_{k,s}-1)+U_{\uparrow \downarrow}\sum_{k}n_{k,\uparrow}n_{k,\downarrow}
\end{equation}
where the index $k$ labels the lattice site, $s \in \{\uparrow, \downarrow\}$ denotes the spin state, $a_{k,s}$ is the annihilation operator, and $n_{k,s} = a_{k,s}^\dagger a_{k,s}$ is the number operator on site $k$. The spin-dependent tunneling amplitude along the lattice direction $\mu$ is denoted by $t_{\mu,s}$, while $U_s$ and $U_{\uparrow\downarrow}$ represent the tunneling and on-site interaction energies, with $\mu = x, y, z$ labeling the lattice directions.

In the single-occupation insulating phase, where $t_{\mu,s}\!\ll\! U_s, U_{\uparrow\downarrow}$ and $\langle n_{i\uparrow}\rangle + \langle n_{i\downarrow}\rangle \approx 1$, the low-energy dynamics reduce to an effective spin Hamiltonian~\cite{duan2003controlling}. By setting $U_s = 2U_{\uparrow\downarrow}$, one obtains the well-known quantum XX model through the Schrieffer–Wolff transformation~\cite{duan2003controlling,schrieffer1966relation,hewson1997kondo}:
\begin{equation}
H=J\sum_{k}^N\left ( \sigma_k^x \sigma_{k+1}^x + \sigma_k^y \sigma_{k+1}^y  \right ) 
\label{eq:XX Hamiltonian}
\end{equation}
where $\sigma_k^x = a_{k\uparrow}^\dagger a_{k\downarrow} + a_{k\downarrow}^\dagger a_{k\uparrow}$, $\sigma_k^y = -i(a_{k\uparrow}^\dagger a_{k\downarrow} - a_{k\downarrow}^\dagger a_{k\uparrow})$, and the effective coupling constant is $J = \tfrac{t_{\mu,\uparrow}t_{\mu,\downarrow}}{U_{\uparrow\downarrow}}$. This mapping shows that the quantum XX Hamiltonian naturally emerges as the low-energy effective model of cold-atom optical lattices, which offers an accessible way to test our theoretical predictions. The eigenstates and eigenenergies of this Hamiltonian are given by \cite{lieb1961two,dutta2015quantum,parkinson2010introduction,amico2008entanglement}
\begin{equation}
\ket{\phi_i}=\sqrt{\frac{2}{N+1}}\sum_{n=1}^N \sin{(\frac{ni\pi}{N+1})}\ket{\bar{1}_n}
\end{equation}

\begin{equation}
E_i=4J\cos{(\frac{i\pi}{N+1})}
\end{equation}
where $i\in\left [1,N  \right ] $ and $\ket{\bar{1}_n}$ is the N-qubit state such that the $\text{n}^{\text{th}}$ qubit (atom) is in the spin up state while all the other qubits are in the spin down states. An important property of this Hamiltonian is that the total z-component of spin $\left \langle \sum_i\sigma_i^Z \right \rangle $ is conserved since $[H,\sigma_{tot}^Z]=0$. This conservation of excitation originates from the underlying $U(1)$ symmetry of the Hamiltonian. Consequently, the Hamiltonian only causes transitions between states with the same total excitation number. For an $N$-qubit system, we consider the initial state of the form:
\begin{equation}
\begin{aligned}
\ket{\psi_0}=&\cos{(\theta)}(a_1\ket{100\cdots00}+a_2\ket{010\cdots00}+\cdots +a_N\ket{000\cdots01})\\
&+e^{i\phi}\sin{(\theta)}\ket{11\cdots11}
\label{eq:ini_state}
\end{aligned}
\end{equation}
where the coefficients $a_i$ are arbitrary complex numbers. The state at time $t$ evolves as
\begin{equation}
\begin{aligned}
\ket{\psi_t}&=\cos{(\theta)}[\sum_{n=1}^Na_n(t)\ket{\bar{1}_n}]+e^{i\phi}\sin{(\theta)}\ket{11\cdots11}\\
&=\cos{(\theta)}[\sum_{k=1}^Nc_k(t)\ket{\phi_k}]+e^{i\phi}\sin{(\theta)}\ket{11\cdots11}\\
\label{eq:t_form}
\end{aligned}
\end{equation}
where the time-dependent coefficients $c_k(t)$ are given by
\begin{equation}
c_k(t)=\sqrt{\frac{2}{N+1}}e^{-iE_kt}[\sum_{n=1}^Na_n\sin{(\frac{nk\pi}{N+1})}]
\end{equation}
In the following, we mainly study the dynamics of different initial states, each characterized by distinct coefficients $a_i$, mixing angle $\theta$, and qubit number $N$.

\begin{figure}[htp]
    \centering
    \includegraphics[width=1\linewidth]{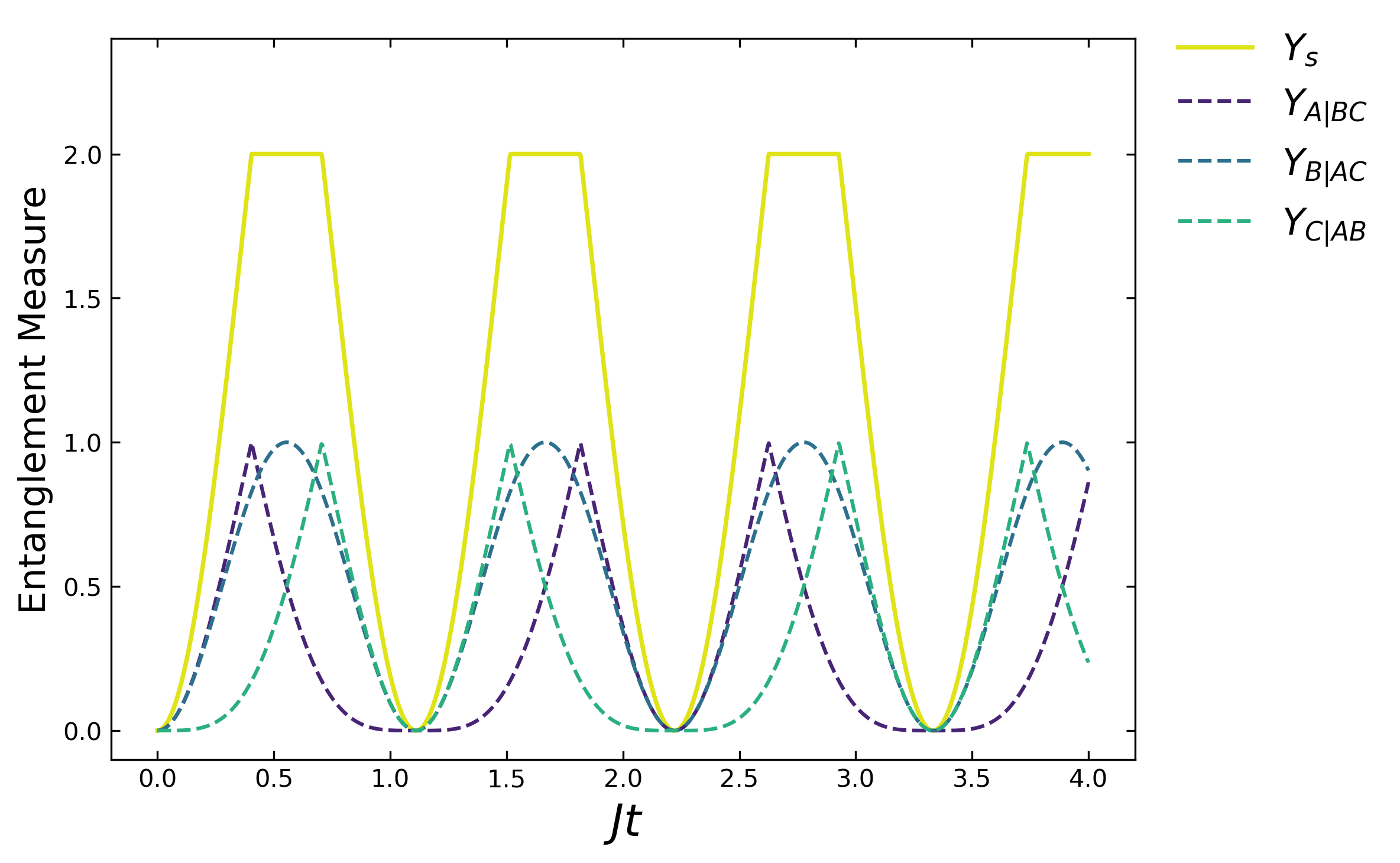}
    \caption{Time evolution of the entanglement volume $Y_s = Y_{A|BC} + Y_{B|AC} + Y_{C|AB}$ for the initial state $\ket{\psi_0} = \ket{001}$. The entanglement volume (solid line) is bounded by $Y_s \le 2$ and displays plateaus where it remains constant. The dashed lines show the individual one-to-other entanglement, which continue to oscillate even when $Y_s$ is frozen. Sharp corners at the plateau boundaries indicate non-analytic transitions in the entanglement dynamics.}
    \label{fig:001}
\end{figure}

\textbf{(i)} We first examine the $N=3$ case with qubits $A$, $B$, and $C$, taking the initial state $\ket{\psi_0} = \ket{001}$, which corresponds to $\theta = 0$ and $\phi = 0$. The entanglement volume $Y_s(t)$, defined by Eq.~\ref{Eq:entanglement volume}, is computed numerically, and the results are shown in Fig.~\ref{fig:001}. The entanglement volume $Y_s$ (solid yellow line) reaches a maximum value of $2$, which represents the upper bound for this dynamics. Notably, $Y_s$ exhibits flat plateaus where it remains constant value at its upper bound for finite time intervals, which is identify as entanglement volume freezing. At the beginning and end of each plateau, $Y_s(t)$ shows sharp corners where its first derivative is discontinuous. The left corner and right corner of each plateaus signal entanglement sudden freezing and sudden thawing, respectively. Meanwhile, the individual one-to-other components $Y_{A|BC}$, $Y_{B|AC}$, and $Y_{C|AB}$ (dashed lines) keep oscillating, indicating that the freezing of $Y_s$ arises from mutual compensation among these terms rather than static entanglement within each bipartition. This same dynamical behavior was also observed in Ref.~\cite{ding2021sudden}, despite differences in the underlying physical model.

\begin{figure}[htp]
    \centering
    \includegraphics[width=1\linewidth]{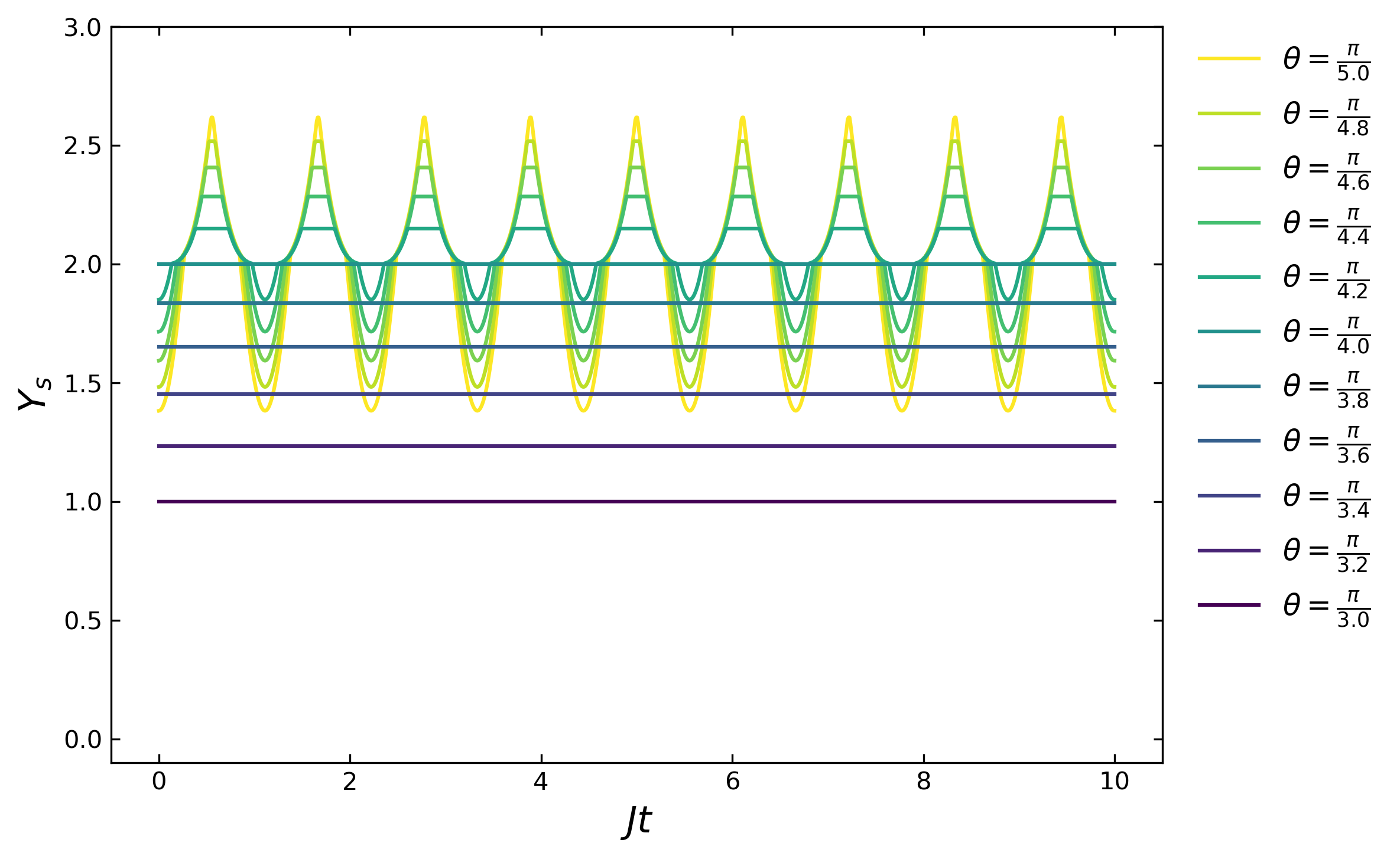}
    \caption{Time evolution of the entanglement volume $Y_s = Y_{A|BC}+Y_{B|AC}+Y_{C|AB}$ for different initial states $\ket{\psi_0} = \cos(\theta)\ket{001} + \sin(\theta)\ket{111}$. For $\theta < \pi/4$, $Y_s$ exhibits periodic freezing plateaus separated by oscillations (temporary freezing). For $\theta \ge \pi/4$, $Y_s$ remains constant in time, indicating permanent freezing.} 
    \label{fig:Ys_theta}
\end{figure}

\textbf{(ii)} We next consider a more general initial state:
\[
\ket{\psi_0} = \cos(\theta)\ket{001} + e^{i\phi}\sin(\theta)\ket{111},
\]
and examine how the parameter $\theta$ affects the entanglement dynamics. The time evolution of the entanglement volume $Y_s(t)$ for various $\theta$ is shown in Fig.~\ref{fig:Ys_theta}. For $\theta < \pi/4$, $Y_s$ exhibits finite-time plateaus, corresponding to temporary entanglement freezing followed by thawing. In contrast, for $\theta \ge \pi/4$, $Y_s$ remains constant throughout the entire evolution, indicating permanent freezing. The boundary $\theta = \pi/4$ thus marks a transition between temporary and permanent freezing regimes. This permanent freezing is nontrivial because the initial state is not an eigenstate of the Hamiltonian. The system continues to evolve dynamically, and each one-to-other Schmidt weight $Y_{A|\bar{A}}$ oscillates in time. However, these oscillations cancel exactly in the sum that defines $Y_s$, keeping the entanglement volume constant.

\begin{figure}[htp]
    \centering
    \includegraphics[width=1\linewidth]{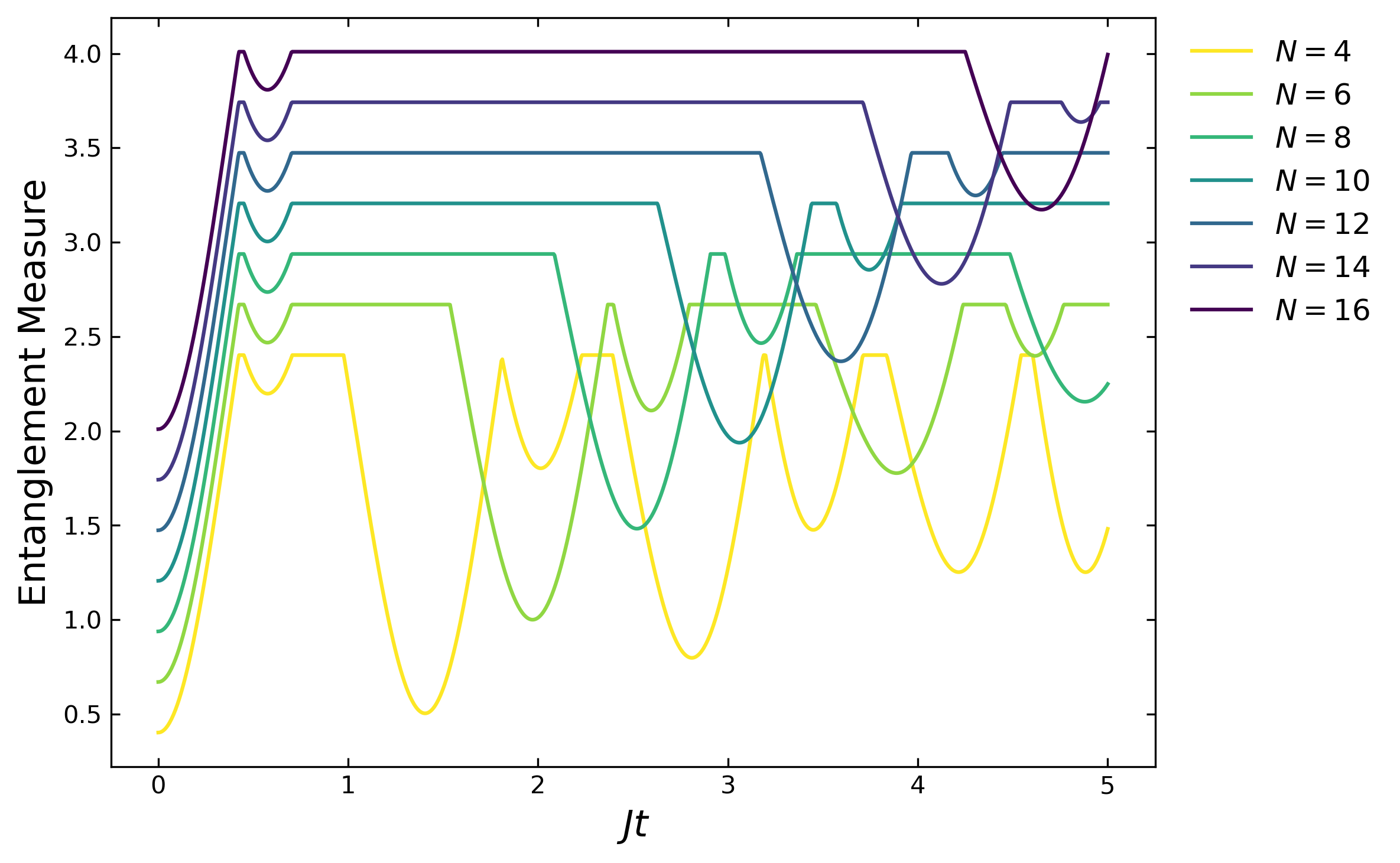}
    \caption{Time evolution of the entanglement volume $Y_s(t)$ for different system sizes $N$ with the initial-state parameter $\theta = \pi/12$. For all values of $N$, $Y_s(t)$ initially increases from zero, passes through a shallow dip, and then enters a broad freezing region. Both the frozen value and the duration of the plateau increase with the qubit number $N$, indicating that larger systems exhibit stronger and more persistent entanglement volume freezing.}
    \label{fig:Ys_Number}
\end{figure}

\textbf{(iii)} We now extend the analysis to the $N$-qubit case, with the initial state 
\[
\ket{\psi_0} = \cos\!\left(\frac{\pi}{12}\right)\ket{10\cdots0} + \sin\!\left(\frac{\pi}{12}\right)\ket{11\cdots1}.
\]
Figure~\ref{fig:Ys_Number} shows the time evolution of the entanglement volume $Y_s(t)$ for different system sizes $N$. As the number of qubits increases, both the frozen value of $Y_s$ and the duration of the freezing plateau increase, indicating that larger systems sustain stronger and longer entanglement volume freezing. This scaling behavior suggests that the freezing phenomenon becomes more robust as the system size grows.

\begin{figure}[htp]
    \centering
    \includegraphics[width=1\linewidth]{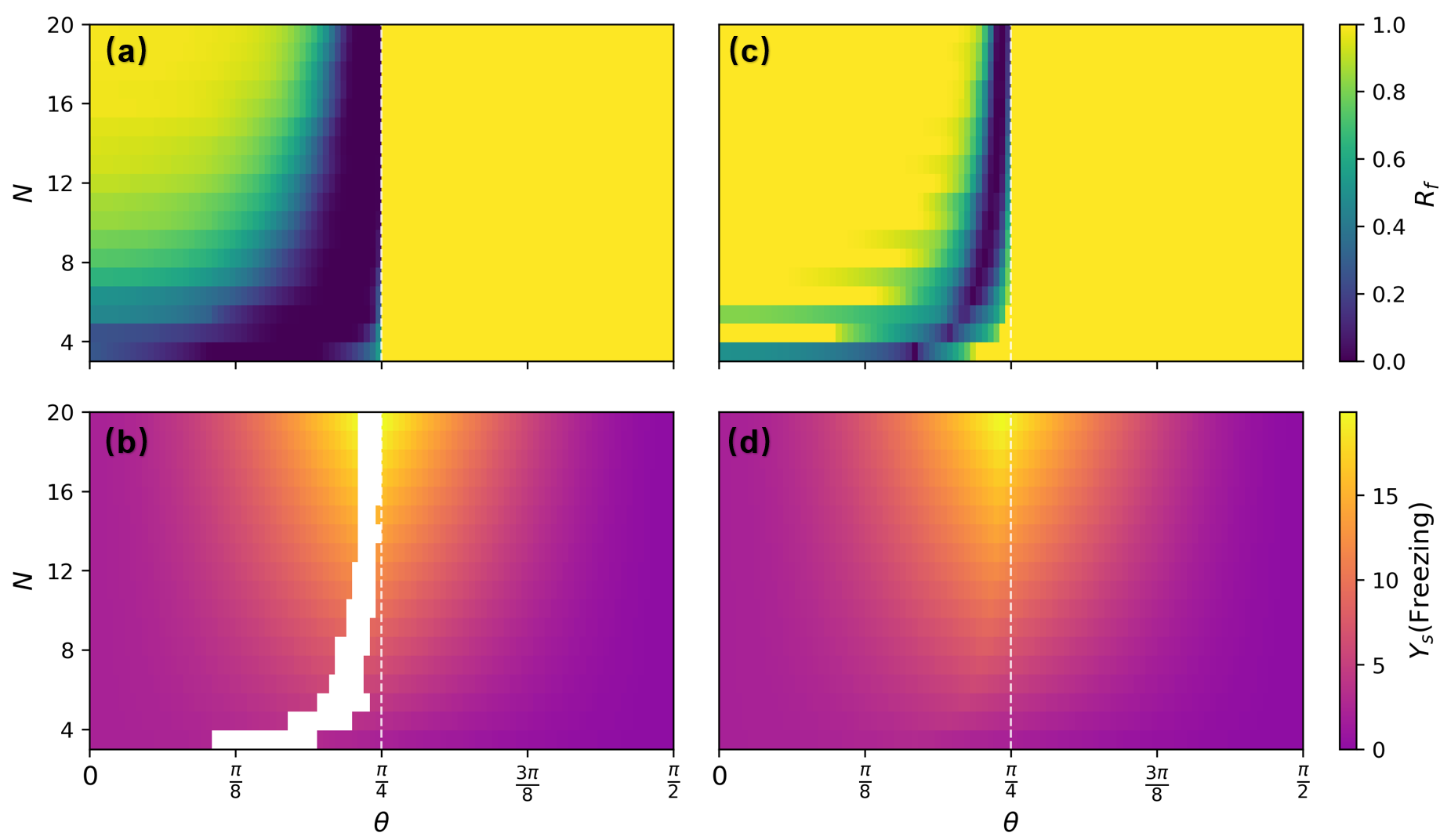}
    \caption{Fraction of freezing time $R_f$ (top row) and frozen entanglement value $Y_s(\mathrm{Freezing})$ (bottom row) 
    as functions of the initial-state parameter $\theta$ and the qubit number $N$. Panels (a,b) correspond to the dynamics of initial state $\ket{\psi_0}=\cos{(\theta)}\ket{10\cdots0}+\sin(\theta)\ket{11\cdots1}$, while panels (c,d) correspond to the dynamics of initial state $\ket{\psi_0}=\frac{\cos{(\theta)}}{\sqrt{3}}(\ket{10\cdots0}+\cdots+\ket{0\cdots010}+\cdots+\ket{0\cdots01})+\sin(\theta)\ket{11\cdots1}$. White regions in (b) mark parameter sets where no freezing occurs.}
    \label{fig:phase_diagram}
\end{figure}

Fig.~\ref{fig:phase_diagram} presents the dynamical phase diagrams of entanglement freezing for different initial states. The horizontal axis represents the initial-state parameter $\theta$, and the vertical axis represents the number of qubits $N$. The upper panels show the fraction of freezing time $R_f$, defined as the ratio between the total duration in which the entanglement volume $Y_s(t)$ remains constant and the total evolution time. The lower panels show the frozen entanglement value $Y_s^{(\mathrm{freeze})}$ as a function of parameters $(N,\theta)$.

In Figs.~\ref{fig:phase_diagram}(a,b), the initial state is chosen as $\cos(\theta)\ket{100\cdots0} + \sin(\theta)\ket{111\cdots1}$. The diagrams clearly exhibit two distinct dynamical regions separated by the boundary $\theta = \pi/4$ (white dashed lines). For $\theta < \pi/4$, the system shows only temporary entanglement freezing, where $R_f$ increases with the system size $N$. Mostly $R_f$ also increases with decreasing $\theta$, but at low N and when $\theta$ is close to $\frac{\pi}{4}$, $R_f$ decreases with decreasing $\theta$. The white regions in Fig.~\ref{fig:phase_diagram}(b) correspond to parameter sets for which no freezing interval is observed. The freezing value increases (if freezing occurs) with increasing $N$ and $\theta$. For $\theta \ge \pi/4$, the freezing becomes permanent. It is also notable that $R_f$ changes abruptly when $\theta$ changes across the boundary $\theta$. The freezing value increases with $N$ and decreases with $\theta$. 

Fig.~\ref{fig:phase_diagram}(c,d) show the results for a symmetrized single-excitation initial state $\ket{\psi_0}=\cos(\theta)(\ket{100}+\ket{010}+\ket{001})/\sqrt{3} + \sin(\theta)\ket{111}$.
The overall structure of the phase diagram remains the same, but $R_f$ is generally larger and exhibits (more obvious) non-monotonic dependence with $N$ and $\theta$. Also, freezing almost always occurs as shown in Fig.~\ref{fig:phase_diagram}(d) since there is no white region. We also note that the freezing value is nearly but not perfectly symmetric around $\theta=\frac{\pi}{4}$.

\subsection{An Open Quantum System}
\label{sec:open}
To further demonstrate the universality of the entanglement volume freezing, we now turn to an open quantum system: two uncoupled cavities interact with two independent $N$-mode optical reservoirs, respectively \cite{ali2014sudden}. The Hamiltonian for each cavity-reservoir pair can be written as
\begin{equation}
\hat{H}=\hbar w \hat{a}^\dagger \hat{a}+\hbar \sum_{k=1}^Nw_k\hat{b}_k^\dagger\hat{b}_k+\hbar\sum_{k=1}^Ng_k(\hat{a}\hat{b}_k^\dagger+\hat{a}^\dagger\hat{b}_k)
\label{eq:openHamil}
\end{equation}

We restrict to the single-excitation regime where each cavity contains at most one photon, and both reservoirs are initially in the vacuum state. If the initial state of the cavity is the one-photon state $\ket{1}_c$ and the initial state of the reservoir is the vacuum state $\ket{\bar{0}}_r=\Pi_{k=1}^N\ket{0_k}_r$, the joint quantum state of each cavity-reservoir pair at any time can be expressed as \cite{ali2014sudden}
\begin{equation}
\begin{aligned}
\ket{\psi_t}_{cr}&=\xi(t)\ket{1}_c\ket{\bar{0}}_r+\sum_{k=1}^N\lambda_k(t)\ket{0}_c\ket{\bar{1}_k}_r\\
&=\xi(t)\ket{1}_c\ket{\bar{0}}_r+\chi(t)\ket{0}_c\ket{\bar{1}}_r
\end{aligned}
\end{equation}
where $\ket{0}_c$ is the vacuum state of the cavity, and $\ket{\bar{1}_k}_r$ denotes the reservoir has one photon in the mode-$k$. The normalized collective single-excitation state of the reservoir is 
$\ket{\bar{1}}_r = \frac{1}{\chi(t)}\sum_{k=1}^N \lambda_k(t)\ket{\bar{1}_k}_r$. Therefore, each cavity-reservoir is an effective two-qubit state. In the limit $N \to \infty$, $\xi(t)$ approaches $e^{-\kappa t / 2}$ and $\chi(t)$ approaches $\sqrt{1 - e^{-\kappa t}}$ \cite{ali2014sudden}.

\begin{figure}[htp]
    \centering
    \includegraphics[width=1\linewidth]{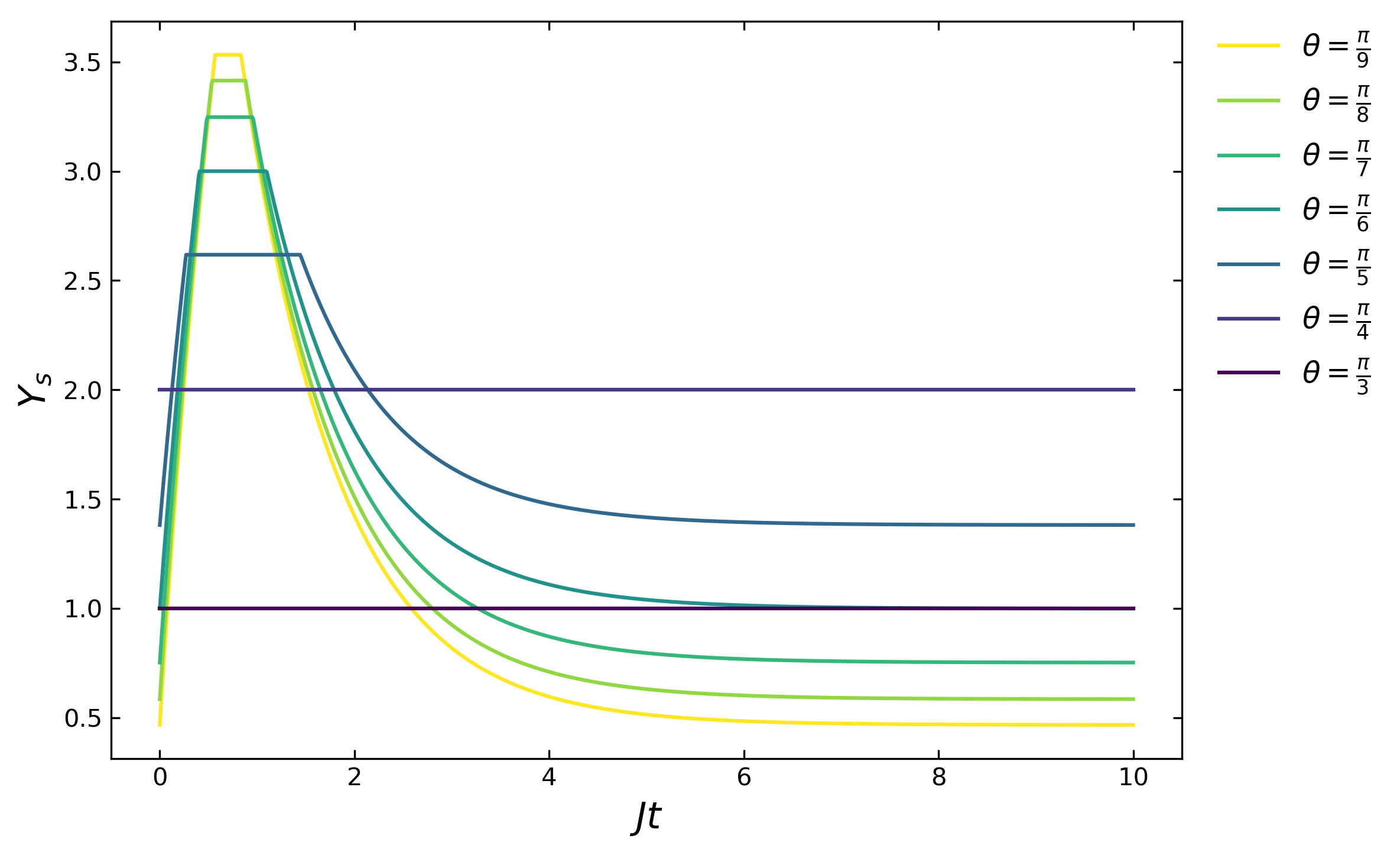}
    \caption{Time evolution of the entanglement volume $Y_s$ for different initial states of the two-cavity system. For $\theta \le \pi/4$, $Y_s(t)$ increases, forms a finite-time freezing plateau, and then decays to a steady value. For $\theta > \pi/4$, $Y_s(t)$ remains constant over the entire evolution, indicating permanent entanglement freezing. Both the frozen value and the freezing duration can be continuously tuned by the initial-state parameter $\theta$.}. 
    \label{fig:opensystem}
\end{figure}

If the initial state of the cavity–cavity pair is 
\[
\ket{\psi_0}_{cc} = \sin(\theta)\ket{00} + \cos(\theta)\ket{11},
\]
the joint state of the two cavities and their corresponding reservoirs at time $t$ is
\begin{equation}
\ket{\psi_t}_{ccrr} = \sin(\theta)\ket{00\bar{0}\bar{0}} 
+ \cos(\theta)\!\left[\chi^2(t)\ket{00\bar{1}\bar{1}} 
+ \xi^2(t)\ket{11\bar{0}\bar{0}} 
+ \chi(t)\xi(t)\big(\ket{10\bar{0}\bar{1}} + \ket{01\bar{1}\bar{0}}\big)\right],
\end{equation}
from which the time-dependent entanglement volume $Y_s(t)$ can be obtained. The results are shown in Fig.~\ref{fig:opensystem}. Because this is an open system, the dynamics lack periodic structure and evolve toward a steady state. For $\theta \le \pi/4$, $Y_s(t)$ first increases, develops a finite-time plateau corresponding to temporary entanglement freezing, and then thaws before gradually decaying to a steady value. For $\theta > \pi/4$, $Y_s(t)$ remains constant throughout the entire evolution, corresponding to permanent freezing. In both regimes, the frozen value and the duration of the freezing interval are tunable through the parameter $\theta$, reflecting direct control of the entanglement dynamics via the initial state.

\begin{figure}[htp]
    \centering
    \includegraphics[width=1\linewidth]{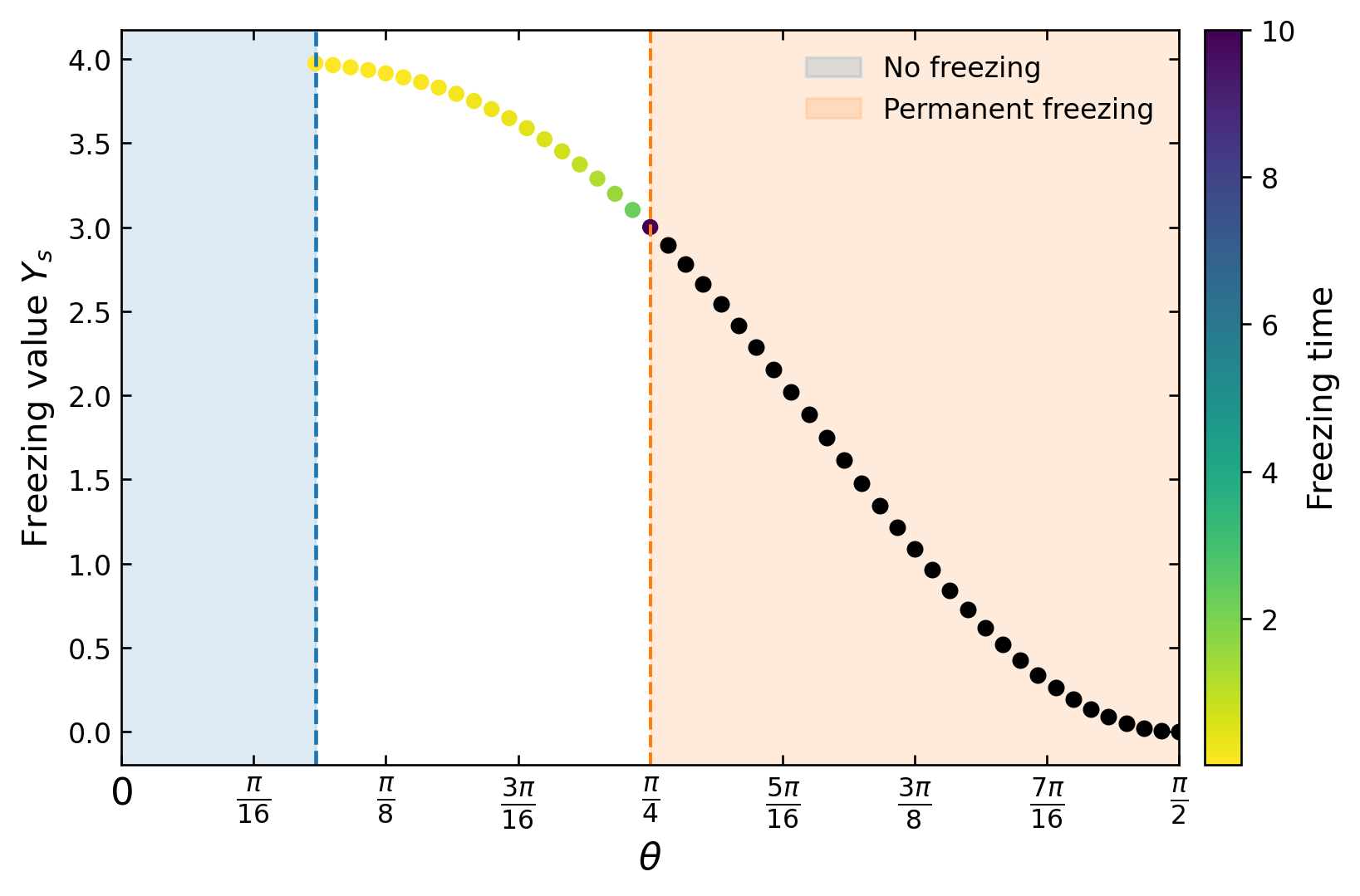}
    \caption{Freezing values versus mixing angle $\theta$ for the same dynamics shown in Fig.~\ref{fig:opensystem}. Each dot shows the freezing value corresponding to the left axis. The color of the dots encodes the freezing time $t_f$. Black dots denote permanent freezing. The light-blue region indicates no freezing regime ($\theta<\theta_{\mathrm{crit}}$), with $\theta_{\mathrm{crit}}$ marked by a dashed line. The light-orange region indicates the permanent freezing regime, separated by the dashed line at $\theta=\pi/4$.}
    \label{fig:3values}
\end{figure}

Fig~\ref{fig:3values} shows how entanglement volume freezing can be controlled via the mixing angle $\theta$. Left axis corresponds to the freezing value while the color of each dot corresponds to the freezing time. Black dots indicate permanent freezing, which occurs for $\theta\ge\frac{\pi}{4}$ and the region is marked by orange. For small value of $\theta$, there exists a critical value $\theta_{\text{crit}}$ that below this value no entanglement volume freezing occurs. In the middle ($\theta_{\text{crit}}\le\theta\le\frac{\pi}{4}$), the freezing value decreases with $\theta$ but the freezing time increases with $\theta$, demonstrating the trade-off between the freezing time and value. 

\section{Universal Origin of Freezing}
\label{section4}
The observation of entanglement freezing in both closed and open systems indicates that this phenomenon does not depend on the specific eigenvalues or eigenstates of a given Hamiltonian, but rather originates from a more fundamental algebraic structure. In both models considered above, the time-dependent state can be written in the general form of
\begin{equation}
\begin{aligned}
\ket{\psi_t} &= \cos(\theta)\big[a_1(t)\ket{100\cdots0} + a_2(t)\ket{010\cdots0} + \cdots + a_N(t)\ket{000\cdots1}\big] \\
&\quad + e^{i\phi}\sin(\theta)\ket{11\cdots1},
\label{eq:ini_state_general}
\end{aligned}
\end{equation}
where $a_i(t)$ are the time-dependent amplitudes. This form is guaranteed by excitation-number conservation, which is a direct consequence of the $U(1)$ symmetry of the Hamiltonians in Eqs.~\ref{eq:XX Hamiltonian} and~\ref{eq:openHamil}. The $U(1)$ symmetry partitions the Hilbert space into invariant subspaces, and the dynamics preserve the excitation number.

Within this representation, the entanglement volume can be expressed as
\begin{equation}
\begin{aligned}
Y_s &= \sum_{i=1}^{N} Y_{A_i|\bar{A}_i} 
   = N - \sum_{i=1}^{N} \big|\,2\cos^2\theta\,|a_i(t)|^2 - \cos(2\theta)\,\big|,
\end{aligned}
\end{equation}
where $a_i(t)$ are the same coefficients appearing in Eq.~\ref{eq:t_form}.  
This expression reveals that the entanglement volume depends on $\theta$ and the magnitudes of the amplitudes $|a_i(t)|$, but not explicitly on their phases or on the microscopic details of the Hamiltonian.  

From this algebraic form, two possible cases can arise:
\begin{itemize}
    \item \textbf{Case 1:} $\quad2\cos^2\theta\,|a_i(t)|^2 - \cos(2\theta)\, \ge 0$ for all $i$.
    \item \textbf{Case 2:} $\quad2\cos^2\theta\,|a_i(t)|^2 - \cos(2\theta)\ \le 0$ for all $i$.
\end{itemize}

Whenever either case is reached, the entanglement volume becomes time-independent, and thus frozen, because the time-dependent amplitudes $a_i(t)$ only appear under the normalization constraint $\sum_i |a_i(t)|^2 = 1$. Substituting this constraint yields constant values of $Y_s$ in each case:
\begin{equation}
\begin{aligned}
\textbf{Case 1:} \quad 
Y_s &= N - 2\!\left(\sum_{i=1}^{N}|a_i(t)|^2\right)\!\cos^2\theta + N\cos(2\theta) \\
&= N - 2\cos^2\theta + N\cos(2\theta) \\
&= 2(N-1)\cos^2\theta,
\label{eq:Ys_freezing1}
\end{aligned}
\end{equation}

\begin{equation}
\begin{aligned}
\textbf{Case 2:} \quad 
Y_s &= N + 2\!\left(\sum_{i=1}^{N}|a_i(t)|^2\right)\!\cos^2\theta - N\cos(2\theta) \\
&= N + 2\cos^2\theta - N\cos(2\theta) \\
&= 2N\sin^2\theta + 2\cos^2\theta.
\label{eq:Ys_freezing2}
\end{aligned}
\end{equation}

Therefore, the freezing of the entanglement volume arises directly from the normalization condition on the amplitudes rather than from any particular spectral property of the Hamiltonian. In both cases, $Y_s$ remains constant at a value determined solely by the initial-state parameter $\theta$ and qubit number $N$. This explains why the phenomenon occurs universally in systems with excitation-number conservation.

\vspace{1ex}
\noindent\textbf{Permanent vs temporary freezing.}  
If $\cos(2\theta) \le 0$, the inequality for Case~1 is always satisfied, leading to permanent freezing. In contrast, when $\cos(2\theta) > 0$, the conditions can only hold for limited time intervals, resulting in temporary freezing followed by entanglement thawing. This explains the transition observed in both systems in Sec.~\ref{section3}, where the boundary between the temporary freezing and permanent freezing is given by $\cos(2\theta) = 0$ or $\theta=\frac{\pi}{4}$. 

\vspace{1ex}
\noindent\textbf{Extension to $e$ excitations.}  
The above reasoning can be extended to initial states containing $e$ excitations:
\begin{equation}
\ket{\psi_0} = \cos\theta (\sum_{p} a_{p}\ket{\text{string}})
+ e^{i\phi}\sin\theta\,\ket{11\cdots11}
\label{k_excitations}
\end{equation}
where $\ket{\text{string}}$ denotes a binary string with $(N-e)$ zeros and $e$ ones, and the index $p$ runs over all possible permutations of such $e$-excitation configurations. The quantity $e$ thus represents the total excitation number. The corresponding frozen values of the entanglement volume are
\begin{subequations}\label{Eq.generalization}
\begin{align}
\textbf{Case 1:} \quad Y_s &= 2(N-e)\cos^2\theta, \\
\textbf{Case 2:} \quad Y_s &= 2N\sin^2\theta + 2e\cos^2\theta.
\end{align}
\end{subequations}
This generalization directly links the excitation number $e$ to the amount of frozen entanglement. Freezing occurs only for $1 \le e \le N-2$, while no freezing is observed for $e = N-1$. The detailed derivation is provided in Appendix~\ref{sec:appA}. For $N \ge 3$, the phenomenon is nontrivial; for $N=2$, freezing arises only for $\ket{\psi_0} = \cos(\theta)\ket{00} + \sin(\theta)\ket{11}$, corresponding to $e = N - 2 = 0$, which represents a trivial limiting case.

\vspace{1ex}
\noindent\textbf{Tunability.} 
A key feature of this freezing phenomenon is its tunability. The parameter $\theta$ determines whether the freezing is temporary or permanent and sets both the frozen value and the duration of the freezing interval. The excitation number $e$ and the total qubit number $N$ provide additional control parameters, as they directly influence the maximum attainable frozen value according to Eq.~\ref{Eq.generalization}. Furthermore, the interaction strength $J$ defines the characteristic time scale of the evolution, allowing one to vary the length of the freezing period without changing its height. Collectively, these parameters render entanglement freezing a highly controllable effect, highlighting its potential for experimental application of quantum technologies.

\vspace{1ex}
\noindent\textbf{Universality.}  
This analysis shows that entanglement freezing arises from the algebraic structure of the quantum state's evolution trajectory. Consequently, it can occur in any system whose dynamics conserves the total excitation number (i.e., with a U(1)-symmetric Hamiltonian). The microscopic details of the system, such as specific eigenstates or eigenvalues, are not the fundamental origin of the phenomenon but affect how it can be tuned or observed. A wide range of physical platforms fall into this category, as summarized in Table~\ref{tab:universality}, highlighting the universality of the phenomenon.
\begin{table}[h]
\centering
\caption{Representative physical systems that could exhibit entanglement volume freezing.}
\label{tab:universality}
\begin{tabular}{l|l}
\hline\hline
\textbf{System} & \textbf{Relevant Conserved Excitation} \\
\hline
Spin-exchange models (XXZ chains, Heisenberg-type systems)& spin-$z$ excitations\\
Cavity-reservoir systems (dissipative QED) & cavity + reservoir modes \\
Jaynes-Cummings / Tavis-Cummings models & atomic + photonic excitations \\
Bose-Hubbard models in Mott regime & fixed particle number sector \\
Coupled harmonic oscillators / phonon modes & phonon modes\\
\hline\hline
\end{tabular}
\end{table}

\section{Discussion}
\label{section5}
Entanglement freezing is typically accompanied by the entanglement sudden change, which has been reported in various quantum systems but lacks a clear and universal explanation. Such sudden changes represent non-analytic points in the entanglement dynamics and call for a more fundamental interpretation. Our analysis suggests a geometric picture in which the key element is the relation between entanglement and geometry of entanglement structure in Hilbert space. Certain invariant region constraint entanglement by enforcing specific algebraic relations among the state amplitudes. During the evolution, a sudden change of entanglement occurs when the quantum state trajectory suddenly breaks into or exit from such a region, where entanglement takes a constant value. Entanglement freezing corresponds to the part of the evolution trajectory that remains entirely within this region. Therefore, the geometry of both the entanglement-invariant subspace and state trajectory in Hilbert space is crucial for understanding the observed freezing and thawing behavior.

\begin{figure}[htp]
\centering
\includegraphics[width=1\linewidth]{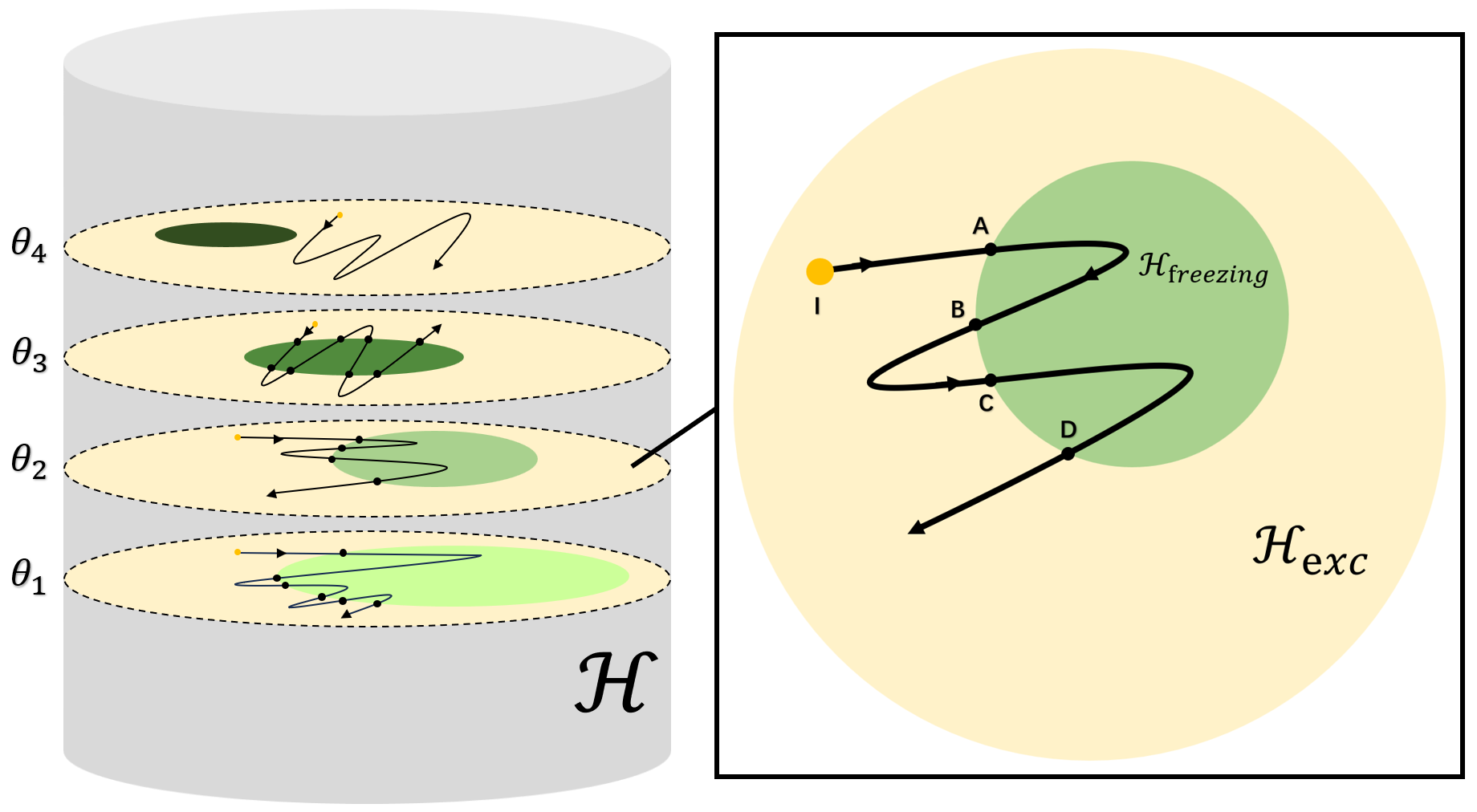}
\caption{Geometric interpretation of entanglement freezing and sudden change.  
The gray cylinder represents the full Hilbert space $\mathcal{H}$, which contains a family of excitation-number-conserving manifolds (yellow layers) characterized by the parameter $\theta$. Within each manifold, a smaller region $\mathcal{H}_{\text{freezing}}$ (green) satisfies stronger algebraic constraints in Eq.~\ref{Eq.generalization} and corresponds to constant entanglement volume. The shade of green indicates the value of the entanglement volume, with darker green regions corresponding to larger values. The black curves illustrates the state trajectory inside one such manifold $\mathcal{H}_{\text{exc}}$. Different values of $\theta$ correspond to distinct excitation manifolds and hence to different freezing regions with their own shape and size.}
\label{fig:geometric dynamics1}
\end{figure}

The Hamiltonians in Eq.~\ref{eq:XX Hamiltonian} and Eq.~\ref{eq:openHamil} both have the feature of excitation-number conservation and thereby restricting the system's evolution to a specific excitation-number-conserving manifold in Hilbert space. Within this manifold, the algebraic relations derived in Sec.~\ref{section3} further define smaller invariant regions. When the quantum state evolves inside this region, the entanglement volume $Y_s$ takes a constant value producing freezing. A sudden change of entanglement corresponds to the trajectory crossing the boundary of this region which produces the non-analytic behavior observed in $Y_s(t)$, as illustrated in Fig.~\ref{fig:geometric dynamics1}. The gray cylinder represents the whole Hilbert space. The yellow area $\mathcal{H}_{\text{exc}}$ denotes excitation-conserving subspace spanned by the initial state with parameter $\theta$. The green subspace $\mathcal{H}_{\text{freezing}}$ represents the smaller entanglement volume invariant region where states obeying the stronger algebraic structure such as either case 1 or case 2 in Eq.~\ref{Eq.generalization}. The shade of green represents different values of frozen entanglement volume. The black line with arrows represents the trajectory of the quantum state dynamics and the arrow denotes the direction of time. Entanglement remains frozen when the trajectory lies inside $\mathcal{H}_{\text{freezing}}$, while points A and C (B and D) mark the sudden entry into (departure from) this region, corresponding to sudden changes of entanglement. Therefore, entanglement freezing happens for the line segments A to B and C to D. 

Fig.~\ref{fig:geometric dynamics1} also reveals the geometric relationship of the entanglement-freezing region. As discussed earlier, this region lies within the excitation-number-conserving manifold characterized by the parameter $\theta$. It is important to note that the excitation-number-conserving manifolds corresponding to different values of $\theta$ do not overlap, owing to the presence of the $\ket{11\cdots1}$ component in the initial state. Consequently, the freezing subspaces $\mathcal{H}_{\text{freezing}}$ associated with different $\theta$ values are also distinct and correspond to different frozen entanglement values. Moreover, $\mathcal{H}_{\text{freezing}}$ for different $\theta$ may vary in both shape and size, leading to variations in the duration of entanglement freezing. By tuning the parameter $\theta$, one effectively changes the slice of the Hilbert space explored by the system’s trajectory. For example, in Fig.~\ref{fig:geometric dynamics1}, the trajectory corresponding to $\theta_4$ does not intersect the freezing region, and hence no freezing occurs for this initial state.

\begin{figure}[htp]
    \centering
    \includegraphics[width=1\linewidth]{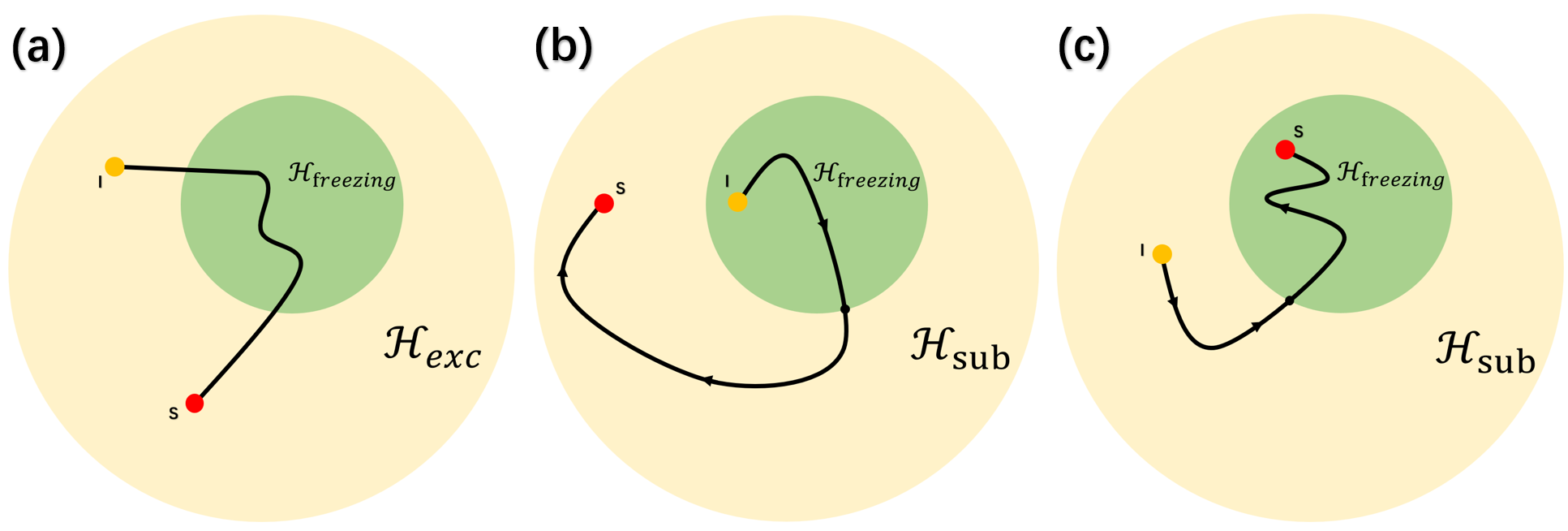}
    \caption{Geometric interpretation of different entanglement-freezing scenarios in open-system dynamics. The yellow area represents a subspace of the full Hilbert space where the quantum state trajectory evolves, while the green area corresponds to an invariant region where the chosen entanglement measure remains constant. The trajectory starts at the initial state (orange dot) and asymptotically approaches a steady state (red dot). (a) Both the initial state and the asymptotic state lie outside the invariant region $\mathcal{H}_{\text{freezing}}$, leading to a finite-time freezing interval in the middle of the evolution. (b) The initial state starts inside the invariant region $\mathcal{H}_{\text{freezing}}$ but the asymptotic state lies outside, resulting in initial freezing that is lost at later times. (c) The initial state begins outside $\mathcal{H}_{\text{freezing}}$ but evolves into it, leading to an initial evolution phase followed by permanent freezing.}
    \label{fig:three cases}
\end{figure}

For the open system dynamics demonstrated in Sec.~\ref{sec:open}, the quantum state trajectory approaches an asymptotic state instead of evolving periodically. Consequently, only one (or finite) freezing interval(s) can be observed before entanglement volume gradually decays to the value determined by the asymptotic state, as shown in Fig.~\ref{fig:three cases} (a), where $\mathcal{H}_{\text{exc}}$ is defined above as a subspace of the whole Hilbert space, the orange dot marks the initial state, and the red dot marks the asymptotic state. 

More generally, our geometric interpretation also applies to a wide range of entanglement freezing phenomenon in open quantum systems \cite{chanda2018scale,mazzola2010sudden,haikka2013non,ali2017freezing,ali2014sudden,wu2016frozen}, as illustrated in Fig.~\ref{fig:three cases} (b,c). The yellow region represents the subspace of the whole Hilbert space where the quantum state trajectory lies, while green area represents any invariant subspace that an entanglement measure of interest takes a constant value. In this work, the yellow region is an excitation-number-conserving manifold and the green region is the entanglement volume invariant subspace. If the initial state starts inside $\mathcal{H}_{\text{sym}}$ but the asymptotic state lies outside [Fig.~\ref{fig:three cases}(b)], one observes temporary freezing followed by thawing. If the initial state starts outside but the asymptotic state lies inside [Fig.~\ref{fig:three cases}(b)], temporary freezing is observed, followed by thawing. Conversely, if the initial state begins outside but the asymptotic state lies inside [Fig.~\ref{fig:three cases}(c)], the system eventually reaches a regime of permanent freezing. This framework also naturally accounts for entanglement sudden death (ESD), which can be regarded as a limiting case of freezing where the invariant subspace corresponds to zero entanglement.

\section{Summary}
\label{section6}
We report an observation of universal sudden freezing and thawing of entanglement volume from the entanglement sharing perspective, arising from excitation number conservation and independent of microscopic details. We reveal a fundamental connection between the non-analytic entanglement dynamics and geometry, providing a geometric explanation in which sudden freezing and thawing correspond to the entry and exit of entanglement invariant subspaces. The entanglement volume freezing effect is highly tunable: the initial mixing angle $\theta$ selects temporary vs.\ permanent freezing and sets both plateau height and duration; the interaction $J$ sets the dynamical time scale without altering the plateau height. In addition, the system size $N$ and excitation number $e$ (or $k$) further determine the attainable frozen value. These parameters together enable precise control over the onset and persistence of freezing.

As the required ingredients, excitation conservation and tunable initial states are naturally available in many experimental platforms, including optical-lattice simulators, cavity/circuit QED setups, trapped ions, and Rydberg arrays, making the predicted freezing dynamics directly accessible to experiments. Our result offers geometry-based insights into the entanglement structure and dynamics in quantum many-body physics, and provide new opportunities for entanglement control in future quantum-engineering applications.

\begin{acknowledgments}
We gratefully acknowledge the late Prof. Joseph H. Eberly, whose insightful discussions and guidance profoundly influenced the development of this work. We also thank Dr. Songbo Xie for valuable comments and assistance with the preparation of the manuscript. This paper is dedicated to the memory of Prof. Eberly.
\end{acknowledgments}

\appendix
\section{Entanglement Volume Freezing for $e$ Excitations}
\label{sec:appA}
In this Appendix, we provide a detailed proof of Eq.~\ref{Eq.generalization} for the general case of $e$ excitations. We also show that entanglement freezing occurs only when the number of excitations is smaller than the total number of qubits minus one, i.e., $e<n-1$.

The joint state at any given time $t$ is given by Eq.~\ref{k_excitations}:
\begin{equation}
\begin{aligned}
\ket{\psi(t)} = &\cos{(\theta)} (a_{1,2,0,\cdots}(t)\ket{110\dots}+a_{1,3,0,\cdots}(t)(t)\ket{1010\dots}+a_{\cdots n-2,n}(t)\ket{0\dots101}\\
&+a_{\cdots n-1,n}(t)\ket{0\dots011}) +\sin{(\theta)}\ket{11\cdots11}\\
=&\cos{(\theta)}\left(\sum_{\text{permutations of}\ \{q_i\}} a_{q_1,q_2,\cdots,q_e} (t)\ket{0,\cdots,1,\cdots,0}\right)+\sin{(\theta)}\ket{11\cdots11}
\end{aligned}
\end{equation}
where the summation runs over all permutations of the indices $\{q_1,q_2,\cdots,q_e\}$ chosen from $\{1,2,\cdots,n\}$. Each index $q_i$ indicates the position of the $i$-th qubit in the excited state $\ket{1}$. 

Because of the permutation symmetry among equivalent excitations, it enough to evaluate the entanglement between one representative qubit (labeled $k$) and the remaining $(n-1)$ qubits. The one-to-other concurrence for this bipartition is
\begin{equation}
C_k = \sqrt{2\bigl(1-\text{Tr}[\rho_k^2]\bigr)},
\end{equation}
where $\rho_k$ is the reduced density matrix of the $k$-th qubit.

Tracing out all qubits except the $k$-th gives
\begin{equation}
\begin{aligned}
&\rho_k=\text{tr}_k[\rho]=\braket{0_k|\psi}\braket{\psi|0_k}+\braket{1_k|\psi}\braket{\psi|1_k}
\end{aligned}
\end{equation}
The two projection components are
\begin{equation}
\begin{aligned}
&\braket{0_k|\psi}=\cos{(\theta)}\sum_{p_1} a_{i_1,i_2,\cdots,i_e}\ket{\phi_k}\\
&\braket{1_k|\psi}=\cos{(\theta)}\sum_{p_2} a_{i_1,\cdots,k,\cdots,i_e}\ket{\chi_k}+\sin{(\theta)}\ket{1,\cdots,1}\\
\label{eq:1phi}
\end{aligned}
\end{equation}
Here $p_1$ denotes all index permutations where $\{i_1,i_2,\cdots,i_e\}$ are selected from $\{1,\cdots,n\}\setminus\{k\}$, and $\ket{\phi_k}$ represents an $(n-1)$-qubit state containing $e$ excitations. Likewise, $p_2$ denotes permutations in which $\{i_1,i_2,\cdots,i_e\}$ are selected from $\{1,\cdots,n\}$ and one of the indices equals $k$, and $\ket{\chi_k}$ is the $(n-1)$-qubit state containing $(e-1)$ excitations. 

The reduced density matrix of the $k$-th qubit can then be expressed as
\begin{equation}
\begin{aligned}
\rho_k =& \cos^2(\theta)\!\left(\sum_{p_1}a_{i_1,i_2,\cdots,i_e}\ket{\phi_k}\right)\!\left(\sum_{p'_1}a^{*}_{i_1,i_2,\cdots,i_e}\bra{\phi_k}\right)\\
&+ \cos^2(\theta)\!\left(\sum_{p_2}a_{i_1,\cdots,k,\cdots,i_e}\ket{\chi_k}\right)\!\left(\sum_{p'_2}a^{*}_{i_1,\cdots,k,\cdots,i_e}\bra{\chi_k}\right)\\
&+ \sin^2(\theta)\ket{1,\cdots,1}\bra{1,\cdots,1}\\
&+ \sin(\theta)\cos(\theta)\!\left[\ket{1,\cdots,1}\!\left(\sum_{p'_2}a^{*}_{i_1,\cdots,k,\cdots,i_e}\bra{\chi_k}\right) + \left(\sum_{p_2}a_{i_1,\cdots,k,\cdots,i_e}\ket{\chi_k}\right)\!\bra{1,\cdots,1}\right].
\label{rho_k}
\end{aligned}
\end{equation}

The condition $e < n-1$ guarantees that $\ket{\chi_k}\neq \ket{1,\cdots,1}$, so the second and third terms in Eq.~\ref{rho_k} remain distinct. If they merged, freezing behavior would vanish.

For clarity, we rewrite $\rho_k$ in matrix form:
\begin{equation}
\rho_k=\begin{pmatrix}
 \cdots &\cdots  &\cdots  &\cdots  &\cdots \\
 \cdots &\mathbf{A}  &\cdots  &\cdots  &\mathbf{c} \\
 \cdots &\cdots  &\mathbf{D} &\cdots  &\cdots \\
 \cdots &\cdots  &\cdots  &\cdots  &\cdots \\
 \cdots &\mathbf{c}^\dagger  &\cdots  &\cdots  &b
\end{pmatrix}
\Rightarrow 
\begin{pmatrix}
 \cdots &\cdots  &\cdots  &\cdots  &\cdots \\
 \cdots &\cdots &\cdots  &\cdots  &\cdots \\
 \cdots &\cdots  &\mathbf{A}  &\cdots  &\mathbf{c} \\
 \cdots &\cdots  &\cdots  &\mathbf{D}  &\cdots \\
 \cdots &\cdots  &\mathbf{c}^\dagger  &\cdots  &b
\end{pmatrix}=
\begin{pmatrix}
 \mathbf{0} &\mathbf{0}   \\
 \mathbf{0} &\mathbf{H}
\end{pmatrix}
\end{equation}
where $\rho_k$ is a $2^n\times2^n$ matrix; $b$ is a scalar element, $\mathbf{c}$ is an $n\times1$ column vector, and $\mathbf{A}$ and $\mathbf{D}$ are square matrices of dimensions $C_n^{k-1}\times C_n^{k-1}$ and $C_n^k\times C_n^k$, respectively. The arrow indicates that, under an appropriate basis reordering, the matrix blocks can be grouped into a smaller Hermitian matrix $\mathbf{H}$ describing the non-zero subspace of $\rho_k$.

The square of $\rho_k$ then takes the form of
\begin{equation}
\rho_k^2=\begin{pmatrix}
 \mathbf{0} &\mathbf{0}   \\
 \mathbf{0} &\mathbf{H}^2
\end{pmatrix}
=
\begin{pmatrix}
 \cdots &\cdots  &\cdots  &\cdots  &\cdots \\
 \cdots &\cdots &\cdots  &\cdots  &\cdots \\
 \cdots &\cdots  &\mathbf{A}^2+\mathbf{c}\mathbf{c}^\dagger  &\cdots  &\mathbf{A}\mathbf{c}+b\mathbf{c} \\
 \cdots &\cdots  &\cdots  &\mathbf{D}^2  &\cdots\\
 \cdots &\cdots  &\mathbf{c}^\dagger\mathbf{A}+b\mathbf{c}^\dagger  &\cdots  &b^2+\mathbf{c}^\dagger\mathbf{c}
\end{pmatrix}
\end{equation}

The trace of $\rho_k^2$ is thus
\begin{equation}
    \text{tr}[\rho_k^2]=\text{tr}[\mathbf{D}^2]+\text{tr}[\mathbf{A}^2+\mathbf{c}\mathbf{c}^\dagger]+b^2+\mathbf{c}^\dagger\mathbf{c}
\end{equation}
where $b=\sin^2(\theta)$. 

The vector $\mathbf{c}$ collects all coefficients associated with $\ket{\chi_k}$ and has an overall amplitude factor $\sin(\theta)\cos(\theta)$:
\begin{equation}
\mathbf{c}=  \sin(\theta)\cos(\theta)\begin{pmatrix}
\vdots \\
 a_{i_1,\cdots,k,\cdots,i_e}\\
\vdots 
\end{pmatrix}
\end{equation}
Therefore,
\begin{equation}
    \text{tr}[\mathbf{c}\mathbf{c}^\dagger]=\text{tr}[\mathbf{c}^\dagger\mathbf{c}]=\sin^2(\theta)\cos^2(\theta)\sum_{p_2} |a_{i_1,\cdots,k,\cdots,i_e}|^2
\end{equation}

Matrices $\mathbf{A}$ and $\mathbf{D}$ represent sub-blocks corresponding to the $\ket{0_k}$ and $\ket{1_k}$ sectors, respectively. They take the forms of
\begin{align}
\mathbf{A} &= \cos^2(\theta)\!\left(\sum_{p_2} a_{i_1,\cdots,k,\cdots,i_e}\ket{\chi_k}\right)\!\left(\sum_{p_2'} a^{*}_{i_1,\cdots,k,\cdots,i_e}\bra{\chi_k}\right),\\
\mathbf{D} &= \cos^2(\theta)\!\left(\sum_{p_1} a_{i_1,i_2,\cdots,i_e}\ket{\phi_k}\right)\!\left(\sum_{p_1'} a^{*}_{i_1,i_2,\cdots,i_e}\bra{\phi_k}\right).
\end{align}

Squaring and taking the trace of each block yields
\begin{align}
\text{Tr}[\mathbf{A}^2] &= \cos^4(\theta)\!\left(\sum_{p_2}|a_{i_1,\cdots,k,\cdots,i_e}|^2\right)^2,\\
\text{Tr}[\mathbf{D}^2] &= \cos^4(\theta)\!\left(\sum_{p_1}|a_{i_1,i_2,\cdots,i_e}|^2\right)^2.
\end{align}

To evaluate $\text{tr}[\rho_k^2]$, the coefficients satisfy the normalization condition:
\begin{equation}
\sum_{p_1}|a_{i_1,i_2,\cdots,i_e}|^2 + \sum_{p_2}|a_{i_1,\cdots,k,\cdots,i_e}|^2 = 1.
\end{equation}
Defining $\sum_{p_2}|a_{i_1,\cdots,k,\cdots,i_e}|^2 \equiv |r_k|^2$ (so that $\sum_{p_1}|a_{i_1,i_2,\cdots,i_e}|^2 = 1-|r_k|^2$), we obtain
\begin{equation}
\begin{aligned}
\text{Tr}[\rho_k^2] &= \cos^4(\theta)(1-|r_k|^2)^2 + \cos^4(\theta)|r_k|^4 + 2\sin^2(\theta)\cos^2(\theta)|r_k|^2 + \sin^4(\theta)\\
&= 2\cos^4(\theta)|r_k|^4 - 2\cos(\theta)\cos(2\theta)|r_k|^2 + \cos^4(\theta) + \sin^4(\theta).
\label{eq:trace_rhok2}
\end{aligned}
\end{equation}

Recalling that $C_k = \sqrt{2(1-\text{Tr}[\rho_k^2])}$ and that the normalized Schmidt weight is defined as $Y_k = 1 - \sqrt{1-C_k^2} = 1 - \sqrt{2\,\text{Tr}[\rho_k^2]-1}$, Eq.~\ref{eq:trace_rhok2} can be simplified as
\begin{equation}
\begin{aligned}
2\,\text{Tr}[\rho_k^2]-1 &= (2\cos^2(\theta)|r_k|-\cos(2\theta))^2.
\end{aligned}
\end{equation}
Thus,
\begin{equation}
Y_k = 1 - \left|\,2\cos^2(\theta)|r_k| - \cos(2\theta)\,\right|.
\end{equation}

Using $\sum_k |r_k|^2 = e$, we can finally compute the entanglement volume $Y_s = \sum_k Y_k$. When $Y_s$ becomes constant in time, the system exhibits entanglement freezing. There are two freezing regimes:
\begin{equation}
Y_s =
\begin{cases}
2(N-e)\cos^2(\theta), & \text{(Case 1)}\\[6pt]
2N\sin^2(\theta) + 2e\cos^2(\theta), & \text{(Case 2)}.
\end{cases}
\end{equation}
These results coincide with Eq.~\ref{Eq.generalization} in the main text and confirm that freezing is independent of the detailed system dynamics.

\nocite{*}

\bibliography{bibliography}

\end{document}